\documentclass[aps,prb,twocolumn,showpacs]{revtex4}

\usepackage{latexsym}
\usepackage{graphicx}
\usepackage{amsmath}
\begin{document}

\title{Strong spin-orbit induced Gilbert damping and g-shift in iron-platinum nanoparticles}
\author{J\"urgen K\"otzler}
\author{Detlef G\"orlitz}
\author{Frank Wiekhorst}
\affiliation{Institut f\"ur Angewandte Physik und Zentrum f\"ur Mikrostrukturforschung, Universit\"at Hamburg, Jungiusstrasse 11, D-20355 Hamburg, Germany}

\date{\today}

\begin{abstract}
The shape of ferromagnetic resonance spectra of highly dispersed, chemically disordered $Fe_{0.2}Pt_{0.8}$ nanospheres is perfectly described by the solution of the Landau-Lifshitz-Gilbert (LLG) equation excluding effects by crystalline anisotropy and superparamagnetic fluctuations. Upon decreasing temperature , the LLG damping $\alpha(T)$ and a negative $g$-shift, $g(T)-g_0$, increase proportional to the particle magnetic moments determined from the Langevin analysis of the magnetization isotherms.  These novel features are explained by the scattering of the $q \to 0$ magnon from an electron-hole (e/h) pair mediated by the spin-orbit coupling, while the sd-exchange can be ruled out. The large saturation values, $\alpha(0)=0.76$ and $g(0)/g_0-1=-0.37$, indicate the dominance of an overdamped 1~meV e/h-pair which seems to originate from the discrete levels of the itinerant electrons in the $d_p=3~nm$ nanoparticles.
\end{abstract}


\maketitle
\section{Introduction}
The ongoing downscaling of magneto-electronic devices maintains the yet intense research of spin dynamics in ferromagnetic structures with restricted dimensions. The effect of surfaces, interfaces, and disorder in ultrathin films \cite{GW05}, multilayers, and nanowires \cite{BH} has been examined and discussed in great detail. On structures confined to the nm-scale in all three dimensions, like ferromagnetic nanoparticles, the impact of anisotropy \cite{SW01} and particle-particle interactions \cite{CD97} on the Ne\'el-Brown type dynamics, which controls the switching of the \textit{longitudinal} magnetization, is now also well understood. On the other hand, the dynamics of the \textit{transverse} magnetization, which e.g. determines the externally induced, ultrafast magnetic switching in ferromagnetic nanoparticles, is still a topical issue. Such fast switching requires a large, i.e. a critical value of the LLG damping parameter $\alpha$ \cite{BH05}. This damping has been studied by conventional \cite{VS81,MR99} and, more recently, by advanced \cite{JS06} ferromagnetic resonance (FMR) techniques, revealing enhanced values of $\alpha$ up to the order of one.  

By now, the LLG damping of \textit{bulk} ferromagnets is almost quantitatively explained by the scattering of the $q=0$ magnon by conduction electron-hole (e/h) pairs due to the spin-orbit coupling $\Omega_{so}$ \cite{VK72}. According to recent \textit{ab initio} bandstructure calculations \cite{JK02}, the rather small values for $\alpha \approx \Omega_{so}^2\tau$ result from the small (Drude) relaxation time $\tau$ of the electrons. For nanoparticles, the Drude scattering and also the wave-vector conservation are ill-defined, and \textit{ab initio} many-body approaches to the spin dynamics should be more appropriate. Numerical work by Cehovin \textit{et al.} \cite{CCM03} considers the modification of the FMR spectrum by the discrete level structure of the itinerant electrons in the particle. However, the effect of the resulting electron-hole excitation, $\epsilon_p \sim v_p^{-1}$, where $v_p$ is the nanoparticle volume on the intrinsic damping has not yet been considered.

Here we present FMR-spectra recorded at $\omega/2\pi$=9.1~GHz on $Fe_{0.2}Pt_{0.8}$ nanospheres, the structural and magnetic properties of which are summarized in Sect.~II. In Sect.~III the measured FMR-shapes will be examined by solutions of the LLG-equation of motion for the particle moments.
Several effects, in particular those predicted for crystalline anisotropy \cite{UN90} and superparamagnetic (SPM) fluctuations of the particle moments \cite{YR92} will be considered. In Sect.~IV, the central results of this study, i.e. the LLG-damping $\alpha(T)$ reaching values of almost one and a large $g$-shift, $g(T)-g_0$, are presented. Since both $\alpha$(T) and $g(T)$ increase proportional to the particle magnetization, they can be related to spin-orbit damping torques, which, due to the large values of $\alpha$ and $\Delta g$ are rather strongly correlated. It will be discussed which features of the e/h-excitations are responsible for these correlations in a nanoparticle. A summary and the conclusions are given in the final section.

\section{Nanoparticle characterization}
The nanoparticle assembly  has been prepared \cite{ES02} following the wet-chemical route by Sun et al. \cite{SS00}. In order to minimize the effect of particle-particle interactions, the nanoparticles were highly dispersed\cite{ES02}. Transmission electron microscopy (TEM) revealed nearly spherical shapes with mean diameter $d_p=3.1 ~nm$ and a rather small width of the log-normal size distribution, $\sigma_d=0.17(3)$. Wide angle X-ray diffraction provided the chemically disordered $fcc$ structure with a lattice constant a$_0$=0.3861~nm \cite{ES02}.

The mean magnetic moments of the nanospheres $\mu_p(T)$ have been extracted from fits of the magnetization isotherms $M(H,T)$, measured by a SQUID magnetometer (QUANTUM DESIGN, MPMS2) in units $emu/g=1.1\cdot10^{20}\mu_B/g$, to  
\begin{equation}
	M_L(H,T)=N_p \mu_p(T)~{\cal{L}}(\frac{\mu_p H}{k_B T}) .
\end{equation}
Here are ${\cal{L}}(y)=\coth(y)-y^{-1}$ with $y=\mu_p(T)H/k_BT$ the Langevin function and $N_p$ the number of nanoparticles per gram. The fits shown in Fig.1(a) demanded for a small paramagnetic background, $M-M_L=\chi_b(T)H$, with a strong Curie-like temperature variation of the susceptibility $\chi_b$, signalizing the presence of paramagnetic impurities. According to the inset of Fig. 1(b) this $1/T$-law turns out to agree with the temperature dependence of the intensity of a weak, narrow magnetic resonance with $g_i=4.3$ depicted in Figs. 2 and 3. Such narrow line with the same $g$-factor has been observed by Berger et al. \cite{BKBB98} on partially crystallized iron-containing borate glass and could be traced to isolated $Fe^{3+}$ ions.

The results for $\mu_p(T)$ depicted in Fig.1(b) show the moments to saturate at $\mu_p(T \to 0)=(910 \pm 30)\mu_B$. This yields a mean moment per atom in the $fcc$ unit cell of $\overline{\mu}(0)=\mu_pa_0^3/4 v_p=0.7\mu_B$ corresponding to a spontaneous magnetization $M_s(0)=5.5~kOe$. According to previous work by Menshikov et al.\cite{AM74} on chemically disordered $Fe_xPt_{1-x}$ this corresponds to an iron-concentration of x=0.20. Upon rising temperature the moments decrease rapidly, which above 40~K can be rather well parameterized by the empirical power law, $\mu_p(T\geq40~K)\sim(1-T/T_C)^{\beta}$ revealing $\beta=2$ and for the Curie temperature $T_C=(320\pm20)~K$. This is consistent with $T_C=(310\pm10)~K$ for $Fe_{0.2}Pt_{0.8}$ emerging from a slight extrapolation of results for $T_C(x\geq0.25)$ of $Fe_xPt_{1-x}$ \cite{IN74}. No quantitative argument is at hand for the exponent $\beta=2$, which is much larger than the mean field value $\beta_{MF}=1/2$. We believe that $\beta=2$ may arise from a reduced thermal stability of the magnetization due to strong fluctuations of the ferromagnetic exchange between $Fe$ and $Pt$ in the disordered structure and also to additional effects of the antiferromagnetic $Fe-Fe$ and $Pt-Pt$ exchange interactions. In this context, it may be interesting to note that for low $Fe$ concentrations, $x \leq 0.3$, bulk $Fe_xPt_{1-x}$ exhibits ferromagnetism only in the disordered structure \cite{AM74}, while structural ordering leads to para- or antiferromagnetism. Recent first-principle calculations of the electronic structure produced clear evidence for the stabilizing effect of disorder on the ferromagnetism in $FePt$ \cite{GB03}.
\begin{figure}[ht]
	\centering
		\includegraphics[width=9cm]{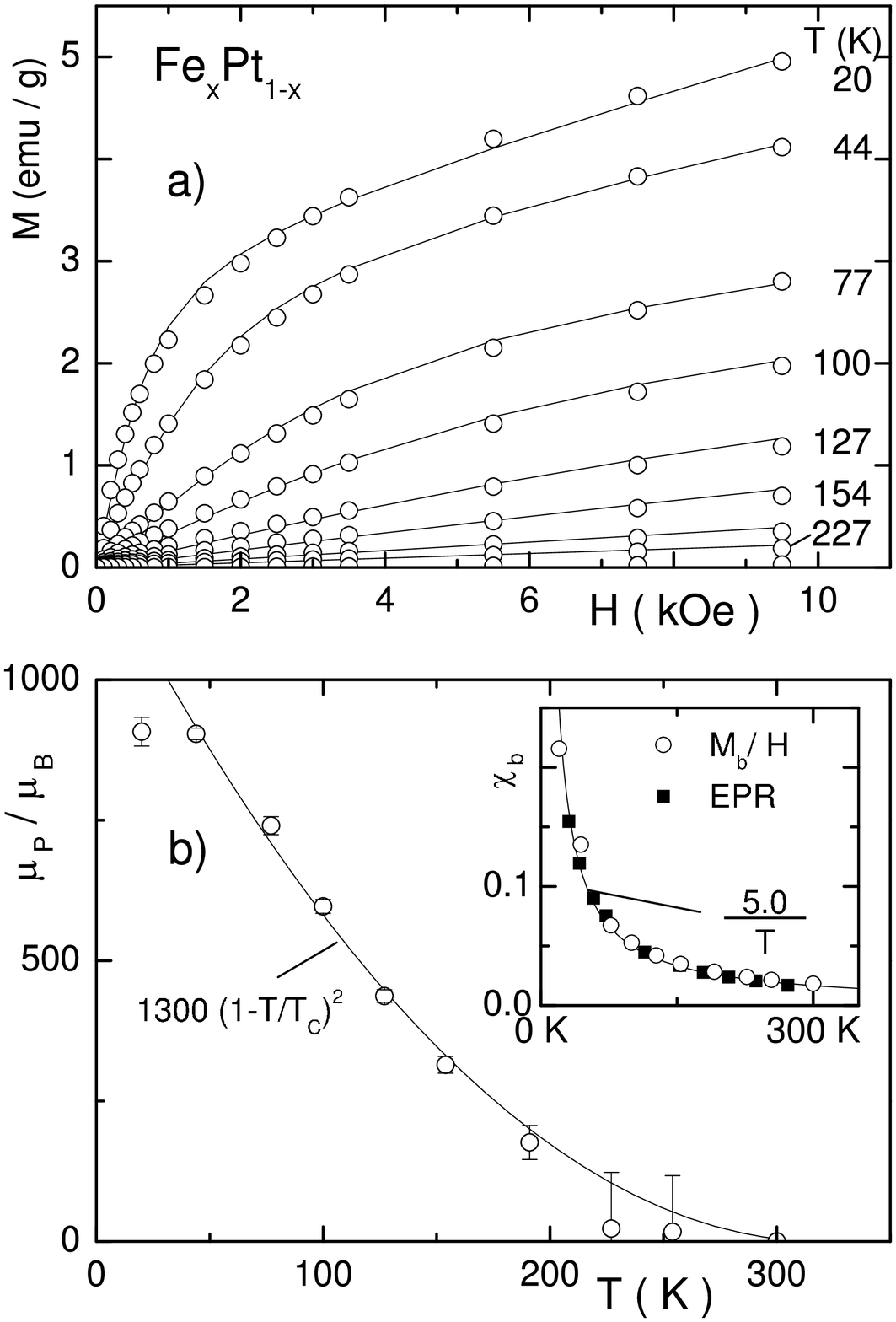}
		\caption{Fig.~1. (a) Magnetization isotherms of the nanospheres fitted to the Langevin model plus a small paramagnetic background $\chi_b \cdot H$; (b) temperature dependences of the magnetic moments of the nanoparticles $\mu_p$ and of the background susceptibility $\chi_b$ ( inset units are emu/g kOe ) fitted to the indicated relations with $T_C=(320 \pm 20)~K$. The inset shows also data from the intensity of the paramagnetic resonance at 1.45 kOe, see Figs.~2 and 3.}
	\label{fig:1}
\end{figure}

From the Langevin fits in Fig.1(a) we obtain for the particle density $N_p=3.5\cdot10^{17}~g^{-1}$. Basing on the well known mass densities of $Fe_{0.2}Pt_{0.8}$ and the organic matrix, we find by a little calculation \cite{FWXX} for the volume concentration of the particles $c_p=0.013$ and, hence, for the mean inter-particle distance, $d_{pp}=d_p/c_p^{1/3}=13.5~nm$. This implies for the maximum (i.e. T=0) dipolar interaction between nearest particles, $\mu_p^2(0)/4\pi\mu_0d_{pp}^3=0.20~K$, so that at the present temperatures, $T \geq 20~K$, the sample should act as an ensemble of independent ferromagnetic nanospheres. Since also the blocking temperature, $T_b=9~K$, as determined from the maximum of the ac-susceptibility at 0.1~Hz in zero magnetic field \cite{FWXX}, turned out to be low, the Langevin-analysis in Fig.1(a) is valid.

\section{Resonance shape}
Magnetic resonances at a fixed X-band frequency of 9.095~GHz have been recorded by a home-made microwave reflectometer equipped with field modulation to enhance the sensitivity. A double-walled quartz tube containing the sample powder has been inserted to a multipurpose, gold-plated VARIAN cavity (model V-4531). Keeping the cavity at room temperature, the sample could be either cooled by means of a continuous flow cryostat (Oxford Instruments, model ESR 900) down to 15 K or heated up to 500~K by an external $Pt$-resistance wire \cite{FWXX}. At all temperatures, the incident microwave power was varied in order to assure the linear response. 

Some examples of the spectra recorded below the Curie temperature are shown in Figs.2 and 3. The spectra have been measured from -9.5~kOe to +9.5~kOe and proved to be independent of the sign of H and free of any hysteresis. This can be expected due to the completely reversible behavior of the magnetization isotherms above 20~K and the low blocking temperature of the particles. Lowering the temperature, we observe a downward shift of the main resonance accompanied by a strong broadening. On the other hand, the position and width of the weak narrow line at $(1.50 \pm0.05)~kOe$ corresponding to $g_i\cong 4.3$ remain independent of temperature. This can be attributed to the previously detected paramagnetic $Fe^{3+}$-impurities \cite{BKBB98} and is supported by the integrated intensity of this impurity resonance $I_i(T)$ evaluated from the amplitude difference of the derivative peaks. Since the intensity of a paramagnetic resonance is given by the paramagnetic susceptibility, $I_i(T)\sim\int{dH~  \chi''_{xx}(H,T) \sim \chi_i(T)}$ can be compared directly to the background susceptibility $\chi_b(T)$, see inset to Fig.1(b). The good agreement between both temperature dependencies suggests to attribute $\chi_b$ to these $Fe^{3+}$-impurities. An analysis of the fitted Curie-constant, $C_i=5~emu K/g~kOe$, yields $N_i=164.10^{17}~g^{-1}$ for the $Fe^{3+}$- concentration, which corresponds to 50 $Fe^{3+}$ per 1150 atoms of a nanosphere.
\begin{figure}[t]
	\centering
		\includegraphics[width=7.5cm]{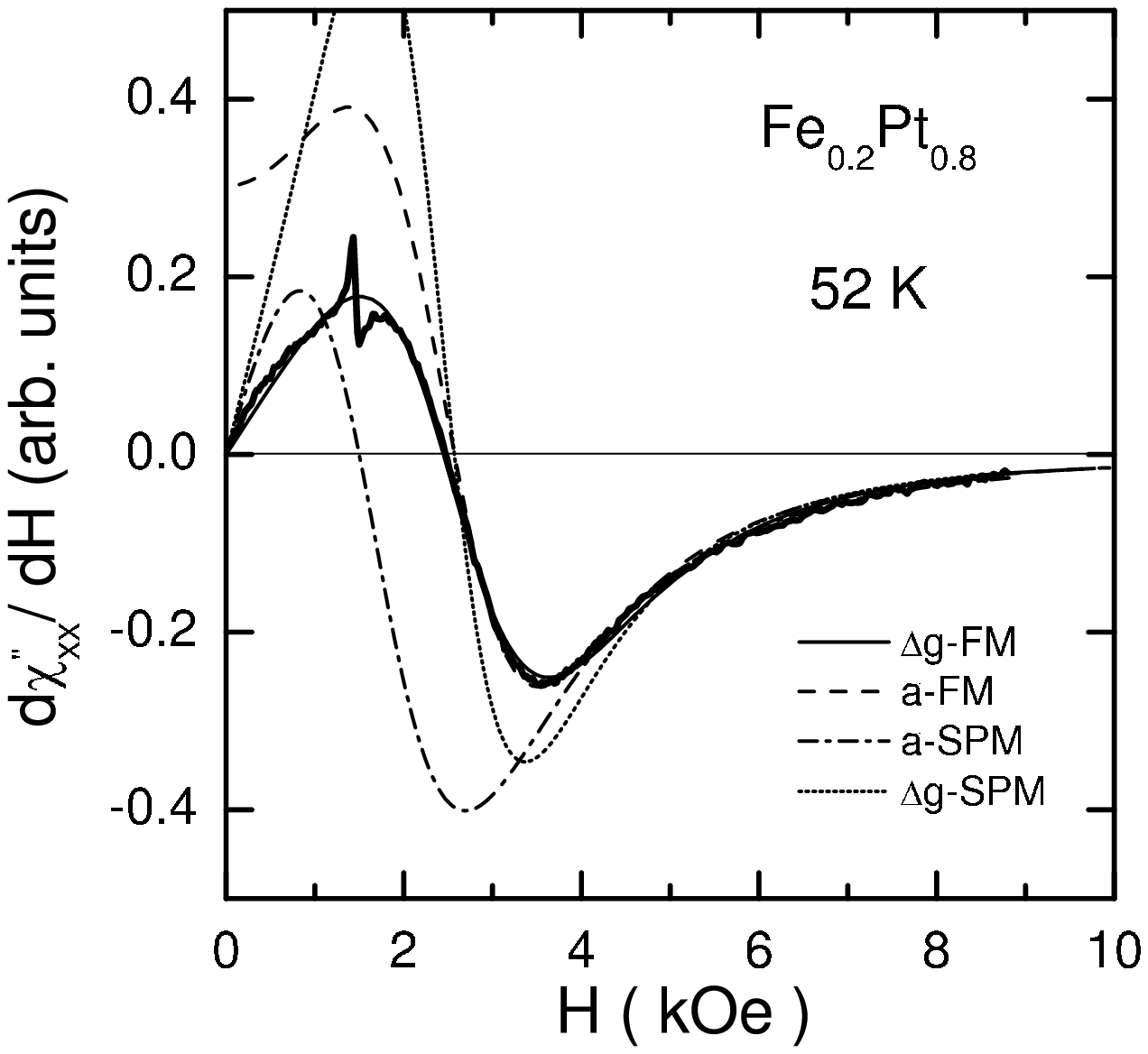}
		\caption{\protect{(a) Derivative of the microwave (f=9.095~GHz) absorption spectrum recorded at T=52~K i.e. close to magnetic saturation of the nanospheres. The solid and dashed curves are based on fits to Eqs.(4) and (8), respectively, which both ignore SPM fluctuations and assume either a $g$-shift and zero anisotropy field $H_A$ ('$\Delta g$-FM') or $\Delta g=0$ and a randomly distributed $H_A$=0.5~kOe ('a-FM'), Eq.~(9). Also shown are fits to predictions by Eq.(11), which account for SPM fluctuations, with $H_A$=0.5~kOe and $\Delta g=0$ ('a-SPM') and, using Eq.(11), for $\Delta g\ne0$ and $H_A$=0 ('$\Delta g$-SPM'). The weak, narrow resonance at 1.45~kOe is attributed to the paramagnetic background with $g=4.3\pm$ 0.1 indicating $Fe^{3+}$~ \cite{BKBB98} impurities.}}
	\label{fig:2}
\end{figure}

\begin{figure}[t]
	\centering
		\includegraphics[width=7.5cm]{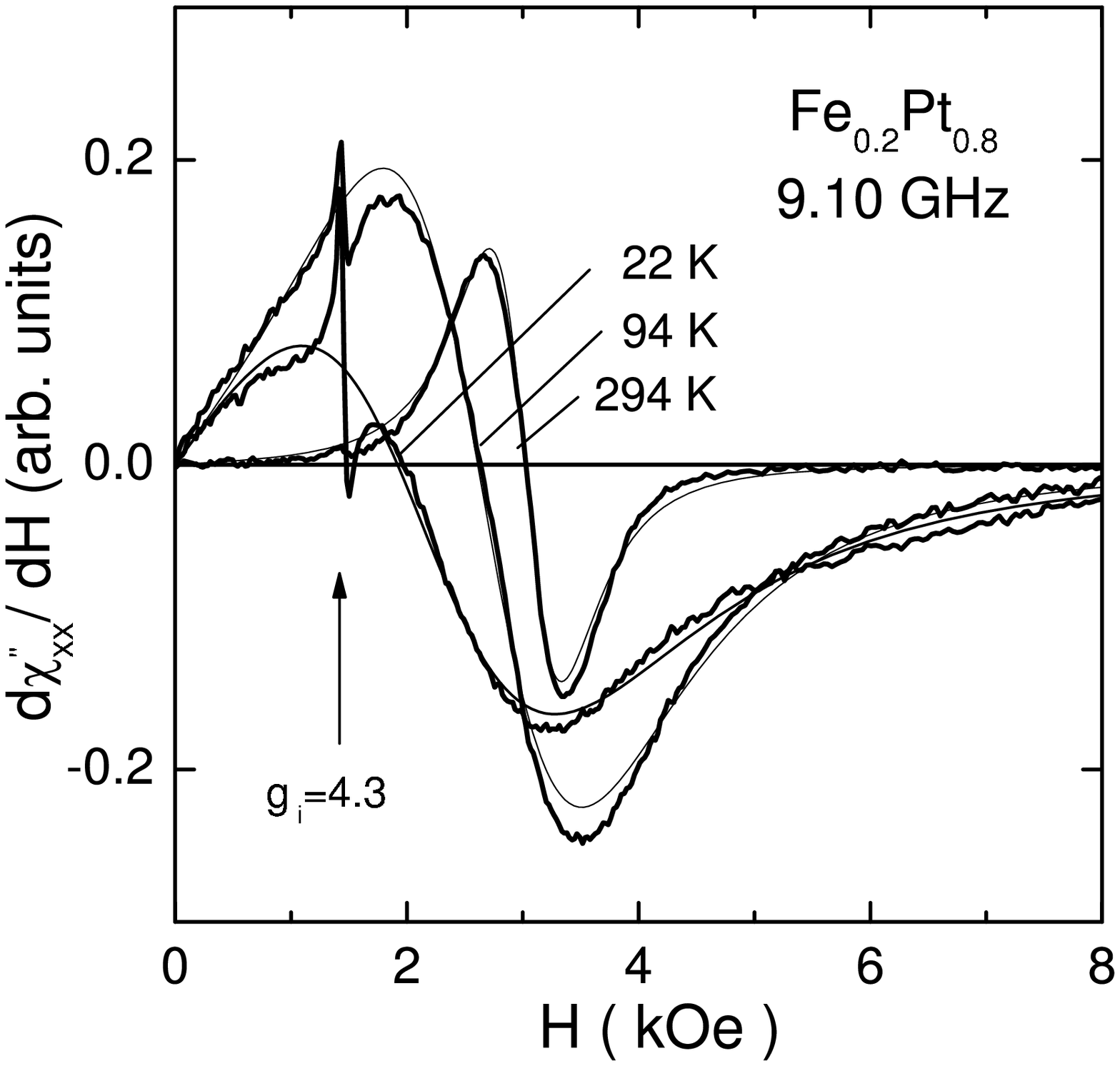}
		\caption{Fig.~3. Derivative spectra at some representative temperatures and fits to Eq.(4). The LLG-damping, $g$-shift, and intensities are depicted in Fig.~4.}
	\label{fig:3}
\end{figure}

With regard to the main intensity, we want to extract a maximum possible information, in particular, on the intrinsic magnetic damping in nanoparticles. Unlike the conventional analysis of resonance fields and linewidths, as applied e.g. to Ni \cite{VS81} and Co \cite{MR99} nanoparticles, our objective is a complete shape analysis in order to disentangle effects by the crystalline anisotropy \cite{UN90}, by SPM fluctuations \cite{YR92}, by an electronic $g$-shift \cite{MU04}, and by different forms of the damping torque $\vec{R}$ \cite{RB00}. Additional difficulties may enter the analysis due to non-spherical particle shapes, size distributions and particle interaction, all of which, however, can be safely excluded for the present nanoparticle assembly.

The starting point of most FMR analyses is the phenomenological equation of motion for a particle moment ( see e.g. Ref. \onlinecite{YR92} )
\begin{equation}
\frac{d}{dt}\vec{\mu}_p=\gamma \vec{H}_{eff} \times\vec{\mu}_p -\vec{R}	\qquad\qquad			,
\label{eom}
\end{equation}
using either the original Landau-Lifshitz (LL) damping with damping frequency $\lambda_L$
\begin{equation}
	\vec{R}_L=\frac{\lambda_L}{M_s} ( \vec{H}_{eff} \times \vec{\mu}_p)\times \vec{s}_p \qquad\qquad,
\end{equation}
or the Gilbert-damping with the Gilbert damping parameter $\alpha_G$,
\begin{equation}
	\vec{R}_G=\alpha_G \frac{d \vec{\mu}_p}{d t} \times \vec{s}_p \qquad\qquad,
\label{RG}
\end{equation}
where $\vec{s}_p=\vec{\mu}_p/\mu_p$ denotes the direction of the particle moment. In Eq.(\ref{eom}), the gyromagnetic ratio, $\gamma=g_0\mu_B/\hbar$, is determined by the regular $g$-factor $g_0$ of the precessing moments . It should perhaps be noted that the validity of the micromagnetic approximation underlying Eq.~(\ref{eom}) has been questioned \cite{NS02} for volumes smaller than $(2\lambda_{sw}(T))^3$, where $\lambda_{sw}=2a_0T_C/T$ is the smallest wavelength of thermally excited spin waves. For the present particles, this estimate leads to a fairly large temperature of $\sim0.7~T_C$ up to which micromagnetics should hold. 

At first, we ignore the anisotropy being small in cubic $Fe_xPt_{1-x}$ \cite{AS94,SSBGOU06}, so that for the present nanospheres the effective field is identical to the applied field, $\vec{H}_{eff}=\vec{H}$. Then, the solutions of Eq.~(\ref{eom}) for the susceptibility of the two normal, i.e. circularly polarized modes, of $N_p$ independent nanoparticles per gram take the simple forms 
\begin{equation}
	\chi_{\pm}^L(H)=N_p \mu_p \gamma~ \frac{1 \mp i \alpha}{\gamma H (1 \mp  i \alpha) \mp \omega}
\label{chiL}
\end{equation}
for $\vec{R}=\vec{R}_L$ with $\alpha=\lambda_L/\gamma M_s$ and for the Gilbert torque $\vec{R}_G$ 
\begin{equation}
	\chi_{\pm}^G(H)=N_p \mu_p \gamma~ \frac{1}{\gamma H\mp \omega(1+ i \alpha_G)}\qquad.
\label{chiG}
\end{equation}
For the LL damping, the experimental, transverse susceptibility, $\chi_{xx}=\frac{1}{2}(\chi_+ + \chi_-)$,  takes the form
\begin{equation}
	\chi_{xx}^L(H)=N_p \mu_p \gamma ~ \frac{\gamma H (1+\alpha^2)-i \alpha \omega}{(\gamma H)^2(1+\alpha^2)-\omega^2-2i \alpha \omega \gamma H}\qquad.
\label{chiexp}
\end{equation}
\\
As the same shape is obtained for the Gilbert torque with $\alpha=\alpha_G$, the damping is frequently denoted as LLG parameter. However, the gyromagnetic ratio in Eq.(\ref{chiexp}) has to be replaced by $\gamma/(1+\alpha^2)$, which only for $\alpha \ll 1$ implies also the same resonance field $H_r$. Upon increasing the damping up to $\alpha\approx 0.7$ (the regime of interest here), the resonance field $H_r$ of $\chi_{xx}^G(H)$ , determined by $d\chi''/dH=0,$ remains constant, $H_r^G\approx \omega/\gamma$, while $H_r^L$decreases rapidly. After renormalization $\gamma/(1+\alpha^2)$ the resonance fields and also the shapes of $\chi^L(H)$ and $\chi^G(H)$ become identical. This effect should be observed when determining the $g$-factor from the resonance fields of broad lines. It becomes even more important if the downward shift of $H_r$ is attributed to anisotropy, as done recently for the rather broad FMR absorption of $Fe_xPt_{1-x}$ nanoparticles with larger $Fe$-content, $x\geq 0.3$ \cite{FW04}. 

In order to check here for both damping torques, we selected the shape measured at a low temperature, T=52~K, where the linewidth proved to be large (see Fig.~2) and the magnetic moment $\mu_p(T)$ was close to saturation (Fig.~1(b)). None of both damping terms could explain both, the observed resonance field $H_r$ and the linewidth $\Delta H=\alpha\omega/\gamma$, and, hence, the lineshape.
By using $\vec{R}_G$, the shift of $H_r$ from $\omega/\gamma=3.00 ~kOe$ was not reproduced by $H_r^G=\omega/\gamma$, while for $\vec{R}_L$ the resonance field $H_r^L$, demanded by the line width, became significantly smaller than $H_r$.

This result suggested to consider as next the effect of a crystalline anisotropy field $\vec{H}_A$ on the transverse susceptibility, which has been calculated by Netzelmann from the free energy of a ferromagnetic grain \cite{UN90}. Specializing his general ansatz to a uniaxial $\vec{H}_A$ oriented at angles $(\theta,\phi)$ with respect to the dc-field $\vec{H}||\vec{e}_z$ and the microwave field, one obtains by minimizing
\begin{eqnarray}
	F(\theta,\phi,\vartheta,\varphi)= -\mu_p [H \cos \vartheta + \hspace*{2cm} \nonumber \\ 
\frac{1}{2} H_A (\sin \vartheta \sin \theta - \cos(\varphi-\phi) +\cos \theta \cos \vartheta)^2]
\label{freeenergy}
\end{eqnarray}
the equilibrium orientation $(\vartheta_0,\varphi_0)$ of the moment $\vec{\mu}_p$ of a spherical grain. After performing the trivial average over $\phi$, one finds for the transverse susceptibility of a particle with orientation $\theta$
\begin{eqnarray}
	\chi_{xx}^L(\theta,H)= \frac{\gamma\mu_p}{2}~\times \hspace*{3cm}     \nonumber \\ 
	\frac{(F_{\vartheta_0 \vartheta_0}+ 
	F_{\varphi_0\varphi_0}/\tan^2\vartheta_0)(1+\alpha^2)-i \alpha\mu_p\omega(1+\cos^2\vartheta_0)}{(1+\alpha^2)(\gamma H_{eff})^2-\omega^2-i \alpha \omega\gamma \Delta H} .\nonumber\\
	\label{chixxL}
\end{eqnarray}
Here $H^2_{eff}=(F_{\vartheta_0 \vartheta_0}F_{\varphi_0\varphi_0}-F^2_{\vartheta_0\varphi_0}/(\mu_p\sin\vartheta_0)^2)$ and $\Delta H=(F_{\vartheta_0 \vartheta_0}+F_{\varphi_0\varphi_0}/\sin^2\vartheta_0)/\mu_p$ are given by the second derivatives of F at the equilibrium orientation of $\vec{\mu}_p$. For the randomly distributed $\vec{H}_A$ of $N_p$ independent particles per gram one has
\begin{equation}
	\chi^L_{xx}(H) = \int^{\pi/2}_{0} d(\cos \theta) \chi^L_{xx}(\theta,H) \qquad.
\label{chiint}
\end{equation}
In a strict sense, this result should be valid at fields larger than the so called thermal fluctuation field $H_T=k_BT/\mu_p(T)$, see e.g. Ref.~\onlinecite{YR92}, which for the present case amounts to $H_T=1.0~kOe$. Hence, in Fig.~2 we fitted the data starting at high fields, reaching there an almost perfect agreement with the curve a-FM. The fit yields a rather small $H_A=0.5~kOe$ which implies a small anisotropy energy per atom, $E_A=\frac{1}{2}\mu_p(0)H_A=1.0~\mu eV$. This number is smaller than the calculated value for bulk $fcc~ FePt$, $E_A=4.0~\mu eV$ \cite{SSBGOU06}, most probably due to the lower $Fe$-concentration (x=0.20) and the strong structural disorder in our nanospheres. We emphasize, that the main defect of this a-FM fit curve arises from the finite value of $d\chi_{xx}''/dH$ at $H=0$. By means of Eq.~(\ref{chixxL}) one finds $\chi_{xx}''(H\to 0,\theta) \sim H_AH/\omega^2$, which remains finite even after averaging over all orientations according to $\theta$ (Eq.~(\ref{chiint})).

The finite value of the derivative of $\chi_{xx}''(H \to 0)$ should disappear if superparamagnetic (SPM) fluctuations of the particles are taken into account. Classical work \cite{JL59} predicted the anisotropy field to be reduced by SPM, $H_A(y)= H_A \cdot (1/L(y)-3/y)$, which for $y=H/H_T\ll 1$ implies $H_A(y)=H_A\cdot y/5$ and, therefore, $\chi_{xx}''(H\to 0)\sim H^2$. A statistical theory for $\chi_{xx}^L(H,T)$ which considers the effect of SPM fluctuations exists only to first order in $H_A/H$ \cite{MU04}. The result of this linear model (LM) in $H_A/H$ which generalizes Eq.~(\ref{RG}), can be cast in the form

\begin{equation}
	\chi^{LM}_{\pm}(\theta,H) = N_p \mu_p L(y) ~\frac{\gamma(1+A\mp i \alpha_A)}{\gamma(1+B \mp i \alpha_B)H\mp \omega}\qquad.
\label{chiLM}
\end{equation}
\\
\noindent
The additional parameters are given in Ref.~\onlinecite{MU04} and contain, depending on the symmetry of $H_A$, higher-order Langevin functions $L_j(y)$ and their derivatives. Observing the validity of the LM for $H \gg H_A=0.5~kOe$, we fitted the data in Fig.~2 to Eq.~\ref{chiLM} with $\chi_{xx}^{LM}(\theta,H)=\frac{1}{2}(\chi_+^{LM}+\chi_-^{LM})$ at larger fields. There one has also $H \gg H_T=1.0~kOe$ and the fit, denoted as a-SPM, agrees with the ferromagnetic result (a-FM). However, increasing deviations appear below fields of $4~kOe$. By varying $H_A$ and $\alpha$, we tried to improve the fit near the resonance $H_r=2.3~kOe$ and obtained unsatisfying results. For low anisotropy, $H_A \leq 3~kOe$, the resonance field could be reproduced only by significantly lower values of $\alpha$, which are inconsistent with the measured width and shape. For $H_A > 3~kOe$, a small shift of $H_r$ occurs, but at the same time the lineshape became distorted, tending to a two-peak structure also found in previous simulations \cite{YR92}. Even at the lowest temperature, $T=22~K$, where the thermal field drops to $H_T=0.4~kOe$, no signatures of such inhomogeneous broadening appear (see Fig.~3). Finally, it should be mentioned that all above attempts to incorporate the anisotropy in the discussion of the lineshape were based on the simplest non-trivial, i.e. uniaxial symmetry, which for $FePt$ was also considered by the theory \cite{SSBGOU06}. For cubic anisotropy, the same qualitative discrepancies were found in our simulations \cite{FWXX}. This insensitivity with respect to the symmetry of $H_A$ originates from the orientational averaging in the range of the $H_A$-values of relevance here.

As a finite anisotropy failed to reproduce $H_r, \Delta H$, and also the shape, we tried a novel ansatz for the magnetic resonance of nanoparticles by introducing a complex LLG parameter,
\begin{equation}
	\hat{\alpha}(T) = \alpha(T)- i~ \beta(T)\qquad .
\label{compa}
\end{equation}
According to Eq.~(\ref{RG}) this is equivalent to a negative $g$-shift, $g(T)-g_0=-\beta(T) g_0$, which is intended to compensate the too large downward shift of $H_r^L$ demanded by $\chi^L_{xx}(H)$ due to the large linewidth. In fact, inserting this ansatz in Eq.~(\ref{chiL}), the fit, denoted as $\Delta g$-FM in Fig.~2, provides a convincing description of the lineshape down to zero magnetic field. It may be interesting to note that the resulting parameters, $\alpha=0.56$ and $\beta=0.27$, revealed the same shape as obtained by using the Gilbert-susceptibilities, Eq.~(\ref{chiG}).

In spite of the agreement of the $\Delta g$-FM model with the data, we also tried to include here SPM fluctuations by using $\hat{\alpha}(T,H)=\hat{\alpha}(T)(1/L(y)-1/y)$ \cite{MU04} for $H_A=0$. The result, designated as $\Delta g$-SPM in Fig.~2 agrees with the $\Delta g$-FM curve for $H \gg H_T$ where $\hat{\alpha}(T,H)=\hat{\alpha}(T)$, but again significant deviations occur at lower fields. They indicate that SPM fluctuations do not play any role here, and this conclusion is also confirmed by the results at higher temperatures. There, the thermal fluctuation field, $H_T=k_BT/\mu_p(T)$, increases to values larger than the maximum measuring field, $H=10~kOe$, so that SPM fluctuations should cause a strong thermal, homogeneous broadening of the resonance due to $\hat{\alpha}(H \gg H_T)=\hat{\alpha} \cdot 2H_T/H$. However, upon increasing temperature, the fitted linewidths, (Fig.~3) and damping parameters (Fig.~4) display the reverse behavior.
\begin{figure}[t]
	\centering
		\includegraphics[width=9cm]{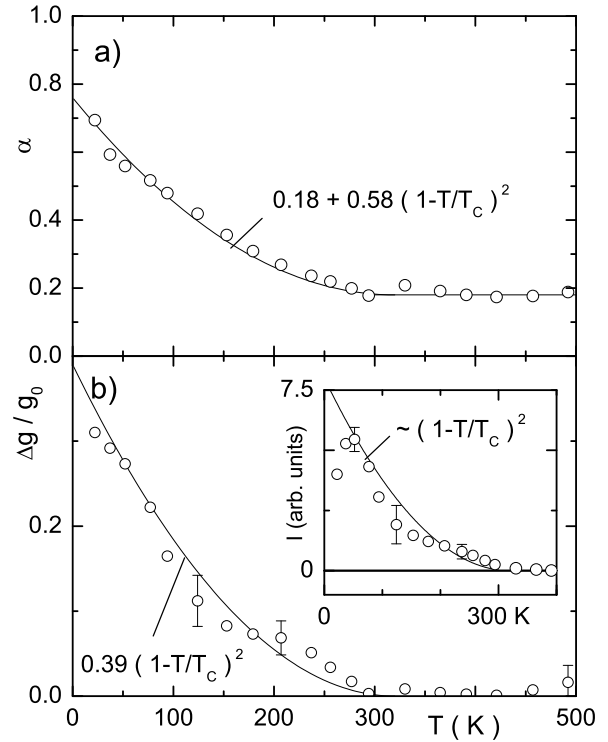}
		\caption{Temperature variation (a) of the LLG-damping $\alpha$ and (b) of the relative $g$-shifts with $g_0=2.16$ (following from the resonance fields at $T >T_C$). Within the error margins, $\alpha(T)$ and $\Delta g(T)$ and also the fitted intensity of the LLG-shape (see inset) display the same temperature dependence as the particle moments in Fig.~1(b).}
	\label{fig:4}
\end{figure}

\section{Complex damping}

In order to shed more light on the magnetization dynamics of the nanospheres we examined the temperature variation of the FMR spectra. Figure~3 shows some examples recorded below the Curie temperature, $T_C=320~K$, together with fits to the $a-$FM model outlined in the last section. Above $T_C$, the resonance fields and the linewidths are temperature independent revealing a mean $g-$factor, $g_0=2.16\pm0.02$, and a damping parameter $\Delta H/H_r=\alpha_0=0.18\pm0.01$. Since $g_0$ is consistent with a recent report on $g$-values of $Fe_xPt_{1-x}$ for $x\geq 0.43$ \cite{MU04}, we suspect that this resonance arises from small $Fe_xPt_{1-x}$-clusters in the inhomogeneous $Fe_{0.2}Pt_{0.8}$ structure. Fluctuations of $g_0$ and of local fields may be responsible for the rather large width. This interpretation is supported by the observation that above $T_C$ the lineshape is closer to a Gaussian than to the Lorentzian following from Eq.(\ref{chiexp}) for small $\alpha$.

The temperature variation for both components of the complex damping, obtained from the fits below $T_C$ to Eq.~(\ref{chiexp}), are shown in Fig.~4. Clearly, they  obey the same power law as the moments, $\mu_p(T)$, displayed in Fig. 1(b), which implies
\begin{equation}
	\hat{\alpha}(T) = (\alpha- i ~\beta)~m_s(T)+\alpha_0\qquad.
\label{hata}
\end{equation}
Here $m_s(T)$=$\mu_p(T)/\mu_p(0)$, $\alpha$=0.58, and $\beta$=-$\Delta g(0)/g_0=0.39$ denote the reduced spontaneous magnetization and the saturation values for the complex damping, respectively. It should be emphasized that the fitted intensity $I(T)$ of the spectra, shown by the inset to Fig.~4(b), exhibits the same temperature variation $I(T)\sim \mu_p(T)$. This behavior is predicted by the ferromagnetic model, Eq.~(\ref{chiexp}), and is a further indication for the absence of SPM effects on the magnetic resonance. If the resonance were dominated by SPM fluctuations, the intensity should decrease like the SPM Curie-susceptibility, $I_{SPM}(T)\sim \mu_p^2(T)/T$, following from Eq.~(\ref{chiLM}), being much stronger than the observed $I(T)$.

At the beginning of a physical discussion of $\hat{\alpha}(T)$, we should point out that the almost perfect fits of the lineshape to Eq.~(\ref{chiexp}) indicate that the complex damping is related to an intrinsic mechanism and that eventual inhomogeneous effects by distributions of particle sizes and shapes in the assembly, as well as by structural disorder are rather unlikely. Since a general theory of the magnetization dynamics in nanoparticles is not yet available, we start with the current knowledge on the LLG-damping in \textit{bulk} and thin film ferromagnets, as recently reviewed by B. Heinrich \cite{BH05}. Based on experimental work on the archetypal metallic ferromagnets and on recent \textit{ab initio} band structure calculations \cite{JK02} there is now rather firm evidence that the damping of the $q$=0-magnon is associated with the torques $\vec{T}_{so}=\vec{m}_s \times \sum_j{(\xi_j\vec{L}_j\times\vec{S})}$ on the spin $\vec{S}$ due to the spin-orbit interaction $\xi_j$ at the lattice sites j. The action of the torque is limited by the finite lifetime $\tau$ of an e/h excitation, the finite energy $\epsilon$ of which may cause a phase, i.e. a $g$-shift. As a result of this magnon - e/h-pair scattering, the temperature dependent part of the LLG damping parameter becomes

\begin{eqnarray}
	\hat{\alpha}(T)-\alpha_0= \frac{\lambda_L(T)}{\gamma M_s(T)}\nonumber\\
	 = \frac{(\Omega_{so}\cdot m_s(T))^2}{\tau^{-1}+i~\epsilon/\hbar}\cdot \frac{1}{\gamma M_s(T)}\qquad .
\label{lambdaL}
\end{eqnarray}
\\
\noindent
For \textit{intraband} scattering, $\epsilon\ll\hbar/\tau$, the aforementioned numerical work \cite{JK02} revealed $\Omega_{so}=0.8\cdot 10^{11}~s^{-1}$ and $0.3\cdot 10^{11}~s^{-1}$ as effective spin-orbit coupling in $fcc~Ni$ and $bcc~Fe$, respectively. Hence, the narrow unshifted ($\Delta g=0)$ \textit{bulk} FMR lines in pure crystals, where $\alpha\leq 10^{-2}$, are related to \textit{intraband} scattering with $\epsilon\ll\hbar/\tau$ and to electronic (momentum) relaxation times $\tau$ smaller than $10^{-13}~s$.

Basing on Eq.~(\ref{lambdaL}), we discuss at first the temperature variation, which implies a linear dependence, $\hat{\alpha}(T)-\alpha_0 \sim m_s(T)$. Obviously,  both, the real and imaginary part of $\hat{\alpha}(T)-\alpha_0$, agree perfectly with the fits to the data in Fig.~4, if the relaxation time $\tau$ remains constant. It may be interesting to note here that the observed temperature variation of the complex damping $\lambda_L(T)$ is not predicted by the classical model \cite{BH67} incorporating the sd-exchange coupling $J_{sd}$. According to this model, which has been advanced recently to ferromagnets with small spin-orbit interaction \cite{YT04} and ferromagnetic multilayers \cite{AC06}, $J_{sd}$ transfers spin from the localized 3d-moments to the delocalized s-electron spins within their spin-flip time $\tau_{sf}$. From the mean field treatment of their equations of motion by Turov \cite{ET61}, we find a form analogous to Eq.~(\ref{lambdaL})

\begin{equation}
	\alpha_{sd}(T)=\frac{\Omega^2_{sd} \chi_s}{\tau_{sf}^{-1}+i \widetilde{\Omega}_{sd}}\cdot \frac{1}{\gamma M_s(T)}
\label{lambdasd}	
\end{equation}
\\
\noindent
where $\Omega_{sd}=J_{sd}/\hbar$ is the exchange frequency , $\chi_s$ the Pauli-susceptibility of the s-electrons and $\widetilde{\Omega}_{sd}/\Omega_{sd}=(1+\Omega_{sd}\chi_s/\gamma M_d)$. The same form follows from more detailed considerations of the involved scattering process (see e.g. Ref.~\onlinecite{BH05}). As a matter of fact, the LLG-damping $\alpha_{sd}=\lambda_{sd}/\gamma M_d$ cannot account for the observed temperature dependence, because $\Omega_{sd}$ and $\chi_s$ are constants. The variation of the spin-torques with the spontaneous magnetization $m_s(T)$ drops out in this model, since the sd-scattering involves transitions between the 3d spin-up and -down bands due to the splitting by the exchange field $J_{sd}~m_s(T)$.

By passing from the bulk to the nanoparticle ferromagnet, we use Eq.~(\ref{lambdaL}) to discuss our results for the complex $\hat{\alpha}(T)$, Eq.~(\ref{hata}). Recently, for $Co$ nanoparticles with diameters 1-4~nm, the existence of a discrete level structure near $\epsilon_F$ has been evidenced \cite{SG99}, which suggests to associate the e/h-energy $\epsilon$ with the level difference $\epsilon_p$ at the Fermi energy. From Eqs.~(\ref{hata}),(\ref{lambdaL}) we obtain relations between $\epsilon$ and the lifetime of the e/h-pair and the experimental parameters $\alpha$ and $\beta$:

\begin{subequations}
\begin{equation}
  \tau^{-1}=\frac{\alpha}{\beta}~\frac{\epsilon}{\hbar}\qquad ,
\label{alpha}
\end{equation}
\begin{equation}
  \frac{\epsilon}{\hbar}=\frac{\beta}{\alpha^2+\beta^2}~\frac{\Omega^2_{so}}{\gamma M_s(0)} \qquad .
\label{beta}
\end{equation}
\end{subequations}
\\
\noindent
Due to $\alpha/\beta=1.5$, Eq.~(\ref{alpha}) reveals a strongly overdamped excitation, which is a rather well-founded conclusion. The evaluation of $\epsilon$, on the  other hand, depends on an estimate for the effective spin-orbit coupling, $\Omega_{so}=\eta_L\chi_e^{1/2}\xi_{so}/\hbar$ where $\eta_L$ represents the matrix element of the orbital angular momentum between the e/h states ~\cite{BH05}. The spin-orbit coupling of the minority $Fe$-spins in $FePt$ has been calculated by Sakuma \cite{AS94}, $\xi_{so}=45~meV$, while the density of states ${\cal{D}}(\epsilon_F)\approx1$/(eV atom) \cite{AS94,KUH96} yields a rather high susceptibility of the electrons, $\eta_L\chi_e=\mu_B^2~{\cal{D}}(\epsilon_F)=4.5\cdot 10^{-5}$. Assuming $\eta_L$=1, both results lead to $\Omega_{so} \approx 3.5\cdot 10^{11}~s^{-1}$, which is by one order of magnitude larger than the values for $Fe$ and $Ni$ mentioned above. One reason for this enhancement and for a large matrix element, $\eta_L$=1, may be the strong hybridization between $3d$ and $4d-Pt$ orbitals \cite{AS94} in $Fe_xPt_{1-x}$. By inserting this result into Eq.~(\ref{beta}) we find $\epsilon=0.8$~ meV. In fact, this value is comparable to an estimate for the level difference at $\epsilon_F$ \cite{SG99} , $\epsilon_p=({\cal{D}}(\epsilon_F)\cdot N_p)^{-1}$ which for our particles with $N_p=(2\pi/3)(d_p/a_0)^3=1060$ atoms yields $\epsilon_p=0.9$~meV.  Regarding the several involved approximations, we believe that this good agreement between the two results on the energy of the e/h excitation, $\epsilon\approx\epsilon_p$, may be accidental. However, we think, that this analysis provides a fairly strong evidence for the magnon-scattering by this excitation, i.e. for the gap in the electronic states due to confinement of the itinerant electrons to the nanoparticle.

\section{Summary And Conclusions}

The analysis of magnetization isotherms explored the mean magnetic moments of $Fe_{0.2}Pt_{0.8}$ nanospheres ($d_p=3.1~nm$) suspended in an organic matrix, their temperature variation up to the Curie temperature $T_C$, the large mean particle-particle distance $D_{pp}\gg d_p$ and the presence of $Fe^{3+}$ impurities. Above $T_C$, the resonance field $H_r$ of the $9.1~GHz$ microwave absorption yielded a temperature independent mean $g$-factor, $g_0=2.16$, consistent with a previous report \cite{MU04} for paramagnetic $Fe_{x}Pt_{1-x}$ clusters. There, the lineshape proved to be closer to a gaussian with rather large linewidth, $\Delta H/H_r=0.18$, which may be associated with fluctuations of $g_0$ and local fields both due to the chemically disordered $fcc$ structure of the nanospheres.

Below the Curie temperature, a detailed discussion of the shape of the magnetic resonance spectra revealed a number of novel and unexpected features.\\ (i) Starting at zero magnetic field, the shapes could be described almost perfectly up to highest field of $10~kOe$ by the solution of the LLG equation of motion for independent ferromagnetic spheres with negligible anisotropy. Signatures of SPM fluctuations on the damping, which have been predicted to occur below the thermal field $H_T=k_BT/\mu_p(T)$, could not be realized.\\ (ii) Upon decreasing temperature, the LLG damping increases proportional to $\mu_p(T)$, i.e. to  the spontaneous magnetization of the particles, reaching a rather large value $\alpha=0.7$ for $T \ll T_C$. We suspect that this high intrinsic damping may be responsible for the absence of the predicted SPM effects on the FMR, since the underlying statistical theory \cite{YR92} has been developed for $\alpha \ll 1$. This conjecture may further be based on the fact that the large intrinsic damping field $\Delta H=\alpha\cdot \omega/\gamma=2.1~kOe$ causes a rapid relaxation of the transverse magnetization ($q=0$ magnon) as compared to the effect of statistical fluctuations of $H_T$ added to $H_{eff}$ in the equation of motion, Eq.(2) \cite{YR92}.\\ (iii) Along with the strong damping, the lineshape analysis revealed a significant reduction of the $g$-factor, which also proved to be proportional to $\mu_p(T)$. Any attempts to account for this shift by introducing uniaxial or cubic anisotropy fields failed, since low values of $\vec{H}_A$ had no effects on the resonance field due to the orientational averaging. On the other hand, larger $\vec{H}_A$'s, by which some small shifts of $H_r$ could be obtained, produced severe distortions of the calculated lineshape.

The central results of this work are the temperature variation and the large magnitudes of both $\alpha(T)$ and $\Delta g(T)$. They were discussed by using the model of the spin-orbit induced scattering of the $q=0$ magnon by an e/h excitation $\epsilon$, well established for bulk ferromagnets, where strong \textit{intraband} scattering with $\epsilon\ll\hbar/\tau$ proved to dominate \cite{BH05}. In nanoparticles, the continuous $\epsilon(\vec{k})$-spectrum of a bulk ferromagnet is expected to be split into discrete levels due to the finite number of lattice sites creating an e/h excitation $\epsilon_p$. According to the measured ratio between damping and $g$-shift, this e/h pair proved to be overdamped, $\hbar/\tau_p=1.5\epsilon_p$. Based on the free electron approximation for $\epsilon_p$ \cite{SG99} and the density of states ${\cal{D}}(\epsilon_F)$ from band-structure calculations for $Fe_xPt_{1-x}$ \cite{AS94,KUH96}, one obtains a rough estimate $\epsilon_p\approx$ 0.9~meV for the present nanoparticles. Using a reasonable estimate of the effective spin-orbit coupling to the minority $Fe$-spins, this value could be well reproduced by the measured LLG damping, $\alpha=0.59$. Therefore we conclude that the novel and unexpected results of the dynamics of the transverse magnetization reported here are due to the presence of a broad e/h excitation with energy $\epsilon_p\approx1~meV$. Deeper quantitative conclusions, however, must await more detailed information on the real electronic structure of nanoparticles near $\epsilon_F$, which are also required to explain the overdamping of the e/h-pairs, as it is inferred from our data.

The authors are indebted to E. Shevchenko and H. Weller (Hamburg) for the synthesis and the structural characterization of the nanoparticles. One of the authors (J.~K.) thanks B.~Heinrich (Burnaby) and M.~F\"ahnle (Stuttgart) for illuminating discussions.

\end{document}